\documentclass[aps,prb,twocolumn,groupedaddress,showpacs,longbibliography]{revtex4-1}
\usepackage{geometry}                		
\usepackage{graphicx}				
\usepackage{amssymb}

\usepackage{graphicx}
\usepackage{natbib}
\usepackage{dcolumn}
\usepackage{bm}
\usepackage{amsmath,amsfonts,amssymb,amsthm,verbatim}
\usepackage{pstricks}
\usepackage[ansinew]{inputenc}
\usepackage{color}
\usepackage{epsfig}

\newcommand{\be}{\begin{equation}}
\newcommand{\ee}{\end{equation}}
\newcommand{\bea}{\begin{eqnarray}}
\newcommand{\eea}{\end{eqnarray}}

\newcommand{\bi}{\begin{itemize}}
\newcommand{\ei}{\end{itemize}}
\newcommand{\bc}{\begin{center}}
\newcommand{\ec}{\end{center}}

\makeatletter

\@ifundefined{textcolor}{}
{%
 \definecolor{BLACK}{gray}{0}
 \definecolor{WHITE}{gray}{1}
 \definecolor{RED}{rgb}{1,0,0}
 \definecolor{GREEN}{rgb}{0,1,0}
 \definecolor{BLUE}{rgb}{0,0,1}
 \definecolor{CYAN}{cmyk}{1,0,0,0}
 \definecolor{MAGENTA}{cmyk}{0,1,0,0}
 \definecolor{YELLOW}{cmyk}{0,0,1,0}
}

\makeatother

\begin{document}
\title{Superconductivity from repulsion: Variational results for the $2D$ Hubbard model in the limit of weak interaction}

\author{Dionys Baeriswyl}
\affiliation{Department of Physics,
             University of Fribourg,\\
             Chemin du Mus\'ee 3, CH-1700 Fribourg, Switzerland}
\affiliation{International Institute of Physics,
             59078-400 Natal-RN, Brazil}
\begin{abstract}
The two-dimensional Hubbard model is studied for small values of the interaction strength ($U$ of the order of the hopping amplitude $t$), using a variational ansatz well suited for this regime. The wave function, a refined Gutzwiller ansatz, has a BCS mean-field state with $d$-wave symmetry as its reference state. Superconducting order is found for densities $n <1$ (but not for $n=1$). This resolves a discrepancy between results obtained with the functional renormalization group, which do predict superconducting order for small values of $U$, and numerical simulations, which did not find superconductivity for $U\lesssim 4t$. Both the gap parameter and the order parameter have a dome-like shape as a function of $n$ with a maximum for $n\approx 0.8$. Expectation values for the energy, the particle number and the superconducting order parameter are calculated using a linked-cluster expansion up to second order in $U$. In this way large systems (millions of sites) can be readily treated and well converged results are obtained. A big size is indeed required to see that the gap becomes very small at half filling and probably tends to zero in the thermodynamic limit, whereas away from half filling a finite asymptotic limit is reached. 
For a lattice of a given size the order parameter becomes finite only above a minimal coupling strength $U_c$. This threshold value decreases steadily with increasing system size, which indicates that superconductivity exists for arbitrarily small $U$ for an infinite system. For moderately large systems the size dependence is quite irregular, due to variations in level spacings at the Fermi energy. The fluctuations die out if the gap parameter spans several level spacings.
\end{abstract}
\maketitle
\section{Introduction}
\label{1}
The Hubbard model was introduced more than fifty years ago
by Anderson \cite{Anderson_59}, Gutzwiller \cite{Gutzwiller_63}, Hubbard \cite{Hubbard_63} and Kanamori \cite{Kanamori_63} for discussing magnetic ordering in narrow-band materials. Later on the model was used for describing the Mott metal-insulator transition \cite{Brinkman_70} and it even served as a microscopic basis for Landau's Fermi Liquid theory of $^3$He\, \cite{Vollhardt_84}. A dramatic upsurge of interest set in when Anderson postulated that the Hubbard model on a square lattice embodies the essentials of cuprate superconductors, by reproducing both the insulating antiferromagnetic phase of the parent compounds and the superconducting phase observed upon doping \cite{Anderson_87}. Still today, it is not clear to what extent Anderson's postulate is corroborated by experiments. 

That superconductivity can arise from purely repulsive interactions was already shown in 1965 by Kohn and Luttinger \cite{Kohn_65}. Using perturbation theory for a continuum model with weak short-range and purely repulsive interaction, they found superconductivity for an unconventional symmetry of the order parameter. During the last thirty years a lot of effort has been spent to find out whether pairing by repulsion is also realized in the two-dimensional Hubbard model, for positive values of the on-site coupling parameter $U$. 

For large values of $U$ there is a general consensus that superconductivity, preferentially with $d$-wave symmetry, exists for a density close to one particle per site, i.e., for a nearly half-filled band \cite{Lee_06, Scalapino_12}. Variational methods have been particularly helpful for estimating the energy gap and the superconducting order parameter, and for studying the competition between antiferromagnetism and superconductivity \cite{Edegger_07, Ogata_08}. Cluster extensions of Dynamical Mean-Field Theory have also found evidence for ($d$-wave) pairing for intermediate to large values of $U$\, \cite{Maier_05}, but different schemes yield rather different results \cite{Rohringer_18}.

For small values of $U$, the method of the functional renormalization group has been particularly helpful for tracking the instabilities of the two-dimensional Hubbard model \cite{Metzner_12}. Initially, superconducting instabilities were detected through divergences of the effective pairing interaction in the normal phase \cite{Zanchi_00, Halboth_00b, Honerkamp_01c, Binz_03}. More recently, techniques were developed to continue the procedure into the superconducting phase, either by using ``partial bosonization'' \cite{Friederich_11} or by combining the scaling in the normal phase with mean-field theory in the ordered phase \cite{Eberlein_14}. While calculations based on the functional renormalization group consistently predict superconductivity ($d$-wave pairing close to half filling), they cannot provide quantitative results for the energy gap or for the order parameter.

Variational calculations do make quantitative predictions, but, besides from being to some extent biased, they usually rely on Monte Carlo sampling, therefore they are limited to modest system sizes (typically $20\times 20$ sites for a square lattice) and suffer from statistical uncertainties. This should not be a problem if the energy gap due to superconductivity is large enough, i.e., for moderate to large values of $U$. However, for small values of $U$, where variational Monte Carlo calculations are unable to find any evidence for superconductivity, one has to worry about the reliability of the method. 

In Gutzwiller's celebrated variational ansatz \cite{Gutzwiller_63} an operator acts on an uncorrelated state to reduce double occupancy. A simple modification refines this ansatz by adding an operator involving the kinetic energy  \cite{Baeriswyl_87, Otsuka_92}. The refined wave function is particularly well suited for treating the small $U$ limit \cite{Otsuka_92, Dzierzawa_95, Tahara_08}. Superconductivity of the two-dimensional Hubbard model has been explored by applying variational Monte Carlo to this ansatz  \cite{Eichenberger_07, Yanagisawa_16, Yanagisawa_19}. 

In the present paper a slightly modified version of the refined Gutzwiller wave function is worked out perturbatively.
Our method is restricted to a relatively small region of coupling strengths, $0.35t\lesssim U\lesssim 1.2t$, where $t$ is the hopping amplitude between nearest-neighbor sites. Nevertheless, the results are clear-cut, especially so because very large system sizes can be treated easily (millions of sites). We choose as ``initial'' uncorrelated state a BCS ansatz with a simple $d$-wave symmetry, where pairing occurs between nearest-neighbor sites. The Hubbard model of course also embodies other orderings, such as antiferromagnetism and $p$-wave superconductivity 
\cite{Hlubina_99}, or superconductivity with more complicated nodal structures than the most simple $p$- and $d$-wave order parameters \cite{Simkovic_16}, but here we limit ourselves to $d$-wave pairing. 

We do find superconductivity
away from half filling (but not at half filling), with a largely increased order parameter as compared to results obtained with the standard Gutzwiller ansatz. The $U$-dependence of the gap parameter is consistent with a power law. This dependence differs from what has been found in RPA-type theories for spin-fluctuation exchange. We also find that superconductivity is not necessarily produced by a lowering of potential energy, but depending on filling it may also be due to a lowering of the ``kinetic energy'' (the expectation value of the hopping term). Therefore non-BCS features are not necessarily a privilege of the large $U$ region of the Hubbard model, i.e., of the doped Mott insulator, but they may also show up already for small values of $U$, i.e., in the ``itinerant part'' of the phase diagram.

The paper is organized as follows. The variational ansatz is introduced in Section \ref{2}, where also the linked-cluster expansion is explained. Section \ref{3} presents the variational ground-state energy for the normal state (vanishing gap parameter), in comparison to a straightforward perturbative expansion in powers of $U$. In Section \ref{4} the formalism is applied to the superconducting state (finite gap parameter). Some details about the minimization procedure are also given. Section \ref{5} deals with the condensation energy and the delicate balance between kinetic- and potential-energy lowering. Results for the gap parameter are discussed in detail in Section \ref{6}, including its ``dome-like'' dependence on particle density as well as an unconventional dependence on $U$. Section \ref{7} introduces the superconducting order parameter and shows that it has a similar dome-like shape as the gap parameter. The dependence on system size is examined in Section \ref{8}.  In  Section \ref{9} the results obtained with the refined wave function are compared with those deduced with the conventional Gutzwiller ansatz. 
The paper is summarized in Section \ref{10}, where also a few problems are listed which could be studied in the future.
Some technicalities are discussed in Appendices \ref{A} and \ref{C}. In Appendix \ref{B} an effective model is treated which has a superconducting mean-field ground state with $d$-wave symmetry.
\section{Variational approach}
\label{2}
\subsection{Hamiltonian}
\label{21}
We consider the (repulsive) Hubbard Hamiltonian $H=H_0+H_{\mbox{\scriptsize int}}$, with nearest-neighbor hopping 
\begin{align}
H_0=-t\sum_{\langle \bf{R,R'}\rangle}\sum_\sigma(c_{{\bf R}\sigma}^\dag c_{{\bf R'}\sigma}^{\phantom{}}+\mbox{h.c.})
\label{eq:hamiltonian}
\end{align}
and on-site repulsion
\begin{align}
H_{\mbox{\scriptsize int}}=U\sum_{\bf R} n_{{\bf R}\uparrow}n_{{\bf R}\downarrow}\, ,
\end{align}
where the operator $c_{{\bf R}\sigma}^\dag$ ($c_{{\bf R}\sigma}^{\phantom{}}$) creates (annihilates) an electron at site ${\bf R}$ with spin projection $\sigma=\uparrow$ or $\downarrow$, and $n_{{\bf R}\sigma}:=c_{{\bf R}\sigma}^\dag c_{{\bf R}\sigma}^{\phantom{}}$. We restrict ourselves to a square lattice with $N_s=L\times L$ sites and a lattice constant 1.
The hopping amplitude is taken as the unit of energy, i.e., we put $t=1$. 
In Fourier space we have
\begin{align}
H_0&=-\sum_{{\bf k}\sigma}\varepsilon_{\bf k}c_{{\bf k}\sigma}^\dag c_{{\bf k}\sigma}^{\phantom{}}\, ,\nonumber\\
H_{\mbox{\scriptsize int}}&=\frac{U}{N_s}\sum_{{\bf k}_1,...,{\bf k}_4}\delta({\bf k}_1-{\bf k}_2+{\bf k}_3-{\bf k}_4)\nonumber\\
&\qquad\qquad\qquad c_{{\bf k}_1\uparrow}^\dag c_{{\bf k}_2\uparrow}^{\phantom{}}c_{{\bf k}_3\downarrow}^\dag 
c_{{\bf k}_4\downarrow}^{\phantom{}}\, ,
\end{align}
where the Kronecker symbol $\delta({\bf k})$ is equal to 1 if ${\bf k}$ is a reciprocal lattice vector and 0 otherwise. 
The tight-binding spectrum is given by
\begin{align}
\varepsilon_{\bf k}=-2\big(\cos k_x+\cos k_y).
\end{align}
The identity
\begin{align}
\delta({\bf k})=\frac{1}{N_s}\sum_{\bf R}e^{i\bf k\cdot R}
\label{eq:Kronecker}
\end{align}
will be extensively used later on.
\subsection{Variational ansatz}
\label{22}
The (standard) Gutzwiller ansatz reads 
\begin{align}
\vert\Psi_G\rangle:=e^{-\eta D}\ \vert\Psi_0\rangle\, ,
\label{eq:gutzwiller}
\end{align}
where $D=\sum_{\bf R}n_{{\bf R}\uparrow}n_{{\bf R}\downarrow}$ measures the number of doubly occupied sites, $\eta$ is a variational parameter and $\vert\Psi_0\rangle$ is the ground state of $H_0$ (the filled Fermi sea). To deal with ordering phenomena, such as antiferromagnetism or superconductivity, one uses, instead of $\vert\Psi_0\rangle$, the ground state $\vert\Psi_m\rangle$ of a symmetry-breaking mean-field Hamiltonian $H_m$ as the reference state. 

The ansatz (\ref{eq:gutzwiller}) has been widely adopted in the limit $\eta\rightarrow\infty$, where double occupancy is completely suppressed and the Hubbard Hamiltonian can be replaced by its large $U$ limit, which corresponds to the $t$-$J$ model \cite{Edegger_07, Ogata_08}. The case of finite $\eta$ with a superconducting reference state has been treated both numerically, using Monte Carlo sampling \cite{Giamarchi_91}, and partly analytically, by a perturbative expansion \cite{Kaczmarczyk_13}. 

Unfortunately, the ansatz (\ref{eq:gutzwiller}) has its weaknesses. Both the energy and the momentum distribution are at odds with exact results in one dimension, where the Gutzwiller wave function can be analyzed exactly \cite{Metzner_87}. 
A remedy was proposed already in the early eighties in terms of an additional prefactor which strengthens the correlations between doubly occupied and empty sites (``doublon-holon binding'') \cite{Kaplan_82}. This additional term turned out to be important for intermediate to large values of $U$, but not for small $U$. Variational Monte Carlo calculations with this modified wave function yield a superconducting ground state in the intermediate to strong coupling regime and for not too large doping, with clear deviations from BCS behavior \cite{Yokoyama_04, Yokoyama_13}. A Jastrow factor producing long-range charge-charge correlations \cite{Yokoyama_90} has also been proposed. It can lead to long-range order in the absence of a symmetry-breaking field \cite{Capello_05, Kaneko_16}. Another way of improving the ansatz (\ref{eq:gutzwiller}) is to modify the reference state $\vert\Psi_m\rangle$. For instance, instead of using two parameters for a superconducting mean-field state (the gap parameter and the ``chemical potential''), one can independently vary the BCS amplitudes $u_{\bf k}$ and in this way introduce 
a huge number of variational parameters (of the order of $N_s$) \cite{Tahara_08}. It has also been proposed to incorporate a ``backflow term'' to improve the accuracy and to account for the kinetic exchange for large values of $U$\, \cite{Tocchio_08}.

Our ansatz reads
\begin{align}
\vert\Psi\rangle:=e^{-\tau H_m}e^{-\eta D}\ \vert\Psi_m\rangle\, ,
\label{eq:ansatz}
\end{align}
where $\vert\Psi_m\rangle$ is the ground state of the mean-field Hamiltonian $H_m$ (with energy eigenvalue $E_m$) and $\tau$
is an additional variational parameter. Eq. (\ref{eq:ansatz}) differs slightly from the ansatz used in previous (variational Monte Carlo) studies \cite{Eichenberger_07, Yanagisawa_16}, where the ``kinetic energy'' $H_0$ was used in the exponent. The choice of $H_m$ instead of $H_0$ is very convenient for a perturbative evaluation of expectation values, as will become clear below. One could even argue that this choice is natural because symmetry breaking is introduced by replacing the eigenstate of $H_0$ by that of $H_m$, therefore it is quite logical to replace also $H_0$ by $H_m$. 

\subsection{Linked cluster expansion}
\label{23}
When Gutzwiller introduced his wave function \cite{Gutzwiller_63}, he adopted the linked cluster expansion for calculating expectation values. A detailed derivation has been given later by Horsch and Fulde \cite{Horsch_79}.
Here we show that the expansion can also be used for our ansatz (\ref{eq:ansatz}).

The variational parameters $\eta,\tau$ plus those defining the mean-field state $\vert\Psi_m\rangle$ are determined by minimizing the energy
\begin{align}
\overline{E}=\frac{\langle\Psi\vert H\vert\Psi\rangle}{\langle\Psi\vert\Psi\rangle}
\end{align}
for a given average number of particles
\begin{align}
\overline{N}=\frac{\langle\Psi\vert N\vert\Psi\rangle}{\langle\Psi\vert\Psi\rangle}\, ,
\label{eq:average_number}
\end{align}
where $N=\sum_{{\bf k}\sigma}c_{{\bf k}\sigma}^\dag c_{{\bf k}\sigma}^{\phantom{}}$.
Eq. (\ref{eq:ansatz}) can be written as
\begin{align}
\vert\Psi\rangle:=e^{-\tau E_m}e^{-\eta D(\tau)}\vert\Psi_m\rangle\, ,
\end{align}
where we have introduced the notation
\begin{align}
{\cal O}(\tau):=e^{-\tau H_m}{\cal O}e^{\tau H_m}
\end{align}
for any operator ${\cal O}$. Correspondingly, we have
\begin{align}
\overline{E}=\frac{\langle\Psi_m\vert e^{-\eta D(-\tau)}He^{-\eta D(\tau)}\vert\Psi_m\rangle}
{\langle\Psi_m\vert e^{-\eta D(-\tau)}e^{-\eta D(\tau)}\vert\Psi_m\rangle}\, ,
\end{align}
and, because of the linked-cluster theorem,
\begin{align}
\overline{E}=\langle\Psi_m\vert e^{-\eta D(-\tau)}He^{-\eta D(\tau)}\vert\Psi_m\rangle_c\, ,
\end{align}
where $c$ means that only those diagrams have to be taken into account where either both exponentials are connected to the operator $H$ or one of the two is connected to $H$ and the two exponential operators are connected to each other. We carry the expansion out to second order in $\eta$ for the hopping term $H_0$ and to first order for the interaction $H_{\mbox{\scriptsize int}}=UD$. This is justified because the optimized correlation parameter $\eta$ is linear in $U$ for $U\rightarrow 0$ and hence the second-order contribution to the interaction part would be proportional to $U^3$, i.e., negligible at this order of the expansion.\footnote{This is strictly valid only in the absence of symmetry breaking, but the argument should remain valid as long as $\Delta_0\ll U$.}
We find
\begin{align}
\overline{E}&\approx\langle H\rangle-2\eta\langle HD(\tau)\rangle_c\nonumber\\
+&\eta^2\left[\langle H_0D^2(\tau)\rangle_c+\langle D(-\tau)H_0D(\tau)\rangle_c\right]\, ,
\label{eq:expansion}
\end{align}
where we have introduced the notation
\begin{align}
\langle{\cal{O}}\rangle:=\langle\Psi_m\vert{\cal{O}}\vert\Psi_m\rangle
\end{align}
for averages with respect to the mean-field ground state. The average particle number is calculated in the same way, and we obtain
\begin{align}
\overline{N}&\approx\langle N\rangle-2\eta\langle ND(\tau)\rangle_c\nonumber\\
+&\eta^2\left[\langle ND^2(\tau)\rangle_c+\langle D(-\tau)ND(\tau)\rangle_c\right]\, .
\label{eq:expansionnumber}
\end{align}
If $\vert\Psi_m\rangle$ is an eigenstate of $N$ all the connected terms vanish and $\overline{N}=\langle N\rangle$. This is the case for 
$\vert\Psi_m\rangle=\vert\Psi_0\rangle$ or for an antiferromagnetic reference state, but not for a BCS state, for which Eq. (\ref{eq:expansionnumber})
together with the constraint of a fixed $\overline{N}$ yields a non-trivial relation between the variational parameters. For $\vert\Psi_m\rangle=\vert\Psi_0\rangle$ the contributions $ \langle H_0D(\tau)\rangle_c$ and $ \langle H_0D^2(\tau)\rangle_c$ also vanish.

For a fixed mean-field state $\vert\Psi_m\rangle$ and a fixed parameter $\tau$, the energy (\ref{eq:expansion}) is easily minimized with respect to $\eta$. Obviously $\eta$ has to be small enough, otherwise the limitation to second order is no longer a good approximation. How small? A simple argument can be given by looking at the problem of two particles on two sites, for which the Gutzwiller ansatz is exact. One readily finds that the second-order expansion corresponds to the replacement
\begin{align}
1-e^{-\eta}\rightarrow \eta-\frac{\eta^2}{2}\, .
\end{align}
For $\eta<0.5$ this approximation is very good, better than $95\%$. This simple estimate agrees with an explicit comparison between a full evaluation of the Gutzwiller ansatz (variational Monte Carlo) and the second-order expansion for the one-dimensional Peierls-Hubbard model, showing good agreement for $\eta\lesssim 0.7$.\cite{Jeckelmann_94}
In the present paper we limit ourselves to the region where $\eta$ does not exceed values of the order of 0.6.  This criterion implies that $U$ should be lower than about 1.2 for the full ansatz
(\ref{eq:ansatz}), while the Gutzwiller wave function (\ref{eq:gutzwiller}) admits $U$-values up to 3.5. 
\section{Normal state}
\label{3}
We discuss first the symmetric case, $H_m=H_0$, $\vert \Psi_m\rangle=\vert \Psi_0\rangle$. The various terms of the expansion (\ref{eq:expansion}) are readily calculated by Wick decomposition. The zeroth-order term is given by
\begin{align}
\langle H\rangle=2\sum_{\bf k}\varepsilon_{\bf k}P_{\bf k}+\frac{N_s}{4}Un^2\, ,
\label{eq:HF}
\end{align}
where 
\begin{align}
P_{\bf k}=\langle c_{{\bf k}\sigma}^\dag c_{{\bf k}\sigma}^{\phantom{}}\rangle
\end{align}
is the Fermi function (which does not depend on $\sigma$ for a non-degenerate state, where all levels are either fully occupied or unoccupied), and 
\begin{align}
n=\frac{\overline{N}}{N_s}=\frac{2}{N_s}\sum_{\bf k}P_{\bf k} 
\end{align}
is the particle density. It is convenient to introduce also the Fermi function of holes,
\begin{align}
Q_{\bf k}=\langle  c_{{\bf k}\sigma}^{\phantom{}}c_{{\bf k}\sigma}^\dag\rangle\, .
\end{align}

\begin{figure}
\begin{center}
\includegraphics[width=7cm]{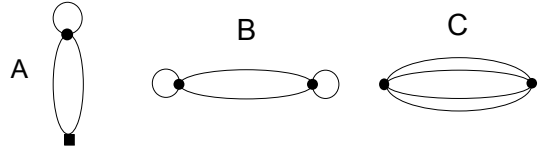}
\end{center}
\caption{First-order diagrams for the expansion (\ref{eq:expansion}). Dots represent two-particle vertices, squares single-particle vertices.}
\label{fig:first_order}
\end{figure}

The first-order contributions are illustrated by diagrams in Fig. \ref{fig:first_order}. The lines represent
\begin{align}
\langle c_{{\bf k}\sigma}^\dag c_{{\bf k}\sigma}^{\phantom{}}(\tau)\rangle=e^{-\tau\vert\xi_{\bf k}\vert}P_{\bf k}
\end{align}
or
\begin{align}
\langle c_{{\bf k}\sigma}^{\phantom{}}c_{{\bf k}\sigma}^\dag (\tau)\rangle=e^{-\tau\vert\xi_{\bf k}\vert}Q_{\bf k}\, ,
\end{align}
where $\xi_{\bf k}=\varepsilon_{\bf k}-\varepsilon_F$ is the single-particle spectrum measured with respect to the Fermi energy $\varepsilon_F$. Thus the parameter $\tau^{-1}$ acts like a soft energy cut-off, which renders correlation effects strongest close to the Fermi surface. We know already that the contribution $A$, which represents the term $\langle H_0D(\tau)\rangle_c$, has to vanish. This follows also from the fact that diagram $A$ involves the factor
 $P_{\bf k}Q_{\bf k}$ $(=0)$. The same is true for the contribution $B$, which therefore also vanishes. To evaluate 
the contribution $C$, we transform the threefold momentum sum into a single sum over lattice sites using Eq. (\ref{eq:Kronecker}). We obtain
\begin{align}
\langle H_{\mbox{\scriptsize int}}D(\tau)\rangle_c=N_sU\sum_{\bf R}\big[P({\bf R},\tau)Q({\bf R},\tau)\big]^2\, ,
\label{eq:1st}
\end{align}
where ($X=P$ or $Q$)
\begin{align}
X({\bf R,\tau}):=\frac{1}{N_s}\sum_{\bf k}e^{i{\bf k\cdot R}}e^{-\tau\vert\xi_{\bf k}\vert}X_{\bf k}\, .
\end{align}

\begin{figure}
\begin{center}
\includegraphics[width=7cm]{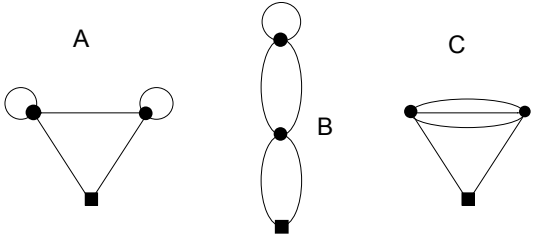}
\end{center}
\caption{Second-order diagrams with symbols as in Fig. \ref{fig:first_order}.}
\label{fig:second_order}
\end{figure}

The second-order contributions are illustrated in Fig. \ref{fig:second_order}. Diagrams $A$ and $B$ yield vanishing contributions, while diagram $C$ can be written as a sum over lattice sites, by defining the quantities
\begin{align}
\varepsilon_1({\bf R},\tau):=\frac{1}{N_s}\sum_{\bf k}e^{i{\bf k\cdot R}}e^{-\tau\vert\xi_{\bf k}\vert}\varepsilon_{\bf k}P_{\bf k}^2\, ,\nonumber\\
\varepsilon_2({\bf R},\tau):=\frac{1}{N_s}\sum_{\bf k}e^{i{\bf k\cdot R}}e^{-\tau\vert\xi_{\bf k}\vert}\varepsilon_{\bf k}Q_{\bf k}^2\, .
\end{align}
We obtain
\begin{align}
&\langle D(-\tau)H_0D(\tau)\rangle_c=
2N_s\sum_{\bf R}P({\bf R},2\tau)Q({\bf R},2\tau)\nonumber\\&\quad\big[\varepsilon_2({\bf R},2\tau)P({\bf R},2\tau)-\varepsilon_1({\bf R},2\tau)Q({\bf R},2\tau)\big]\, .
\label{eq:2nd}
\end{align}

The minimization of Eq. (\ref{eq:expansion}) with respect to $\eta$ yields the correlation energy
\begin{align}
E_{\mbox{\scriptsize corr}}&:=\overline{E}-\langle H\rangle\nonumber\\&
\approx -U^2\frac{\langle DD(\tau)\rangle_c^2}{\langle D(-\tau)H_0D(\tau)\rangle_c}\, ,
\end{align}
which is readily evaluated numerically because the individual terms, Eqs. (\ref{eq:1st}) and (\ref{eq:2nd}), are simple sums. 
The result has still to be minimized with respect to the parameter $\tau$. It turns out that $\tau$ depends weakly on $U$ (for small $U$), but more sensitively on the particle density $n$. Some results for the particular case of $U=1$ are given in Table \ref{table:normal} for a $1000\times 1000$ square lattice. Both $\eta$ and $\tau$ are largest at half filling. 
\begin{table}
\begin{center}
\begin{tabular}{lllll}
$\, n$&\qquad $\eta$&\qquad $\tau$& $\quad E_{\mbox{\scriptsize corr}}/L$&$\quad E_{\mbox{\scriptsize corr}}^{(\mbox{\scriptsize ex})}/L$\\
\hline
1.0&\, 0.59633&\, 0.19805&\, -0.012169&\, -0.012562\\
0.9&\, 0.57690&\, 0.19286&\, -0.011774&\, -0.012072\\
0.8&\, 0.54594&\, 0.18435&\, -0.010787&\, -0.010995\\
0.7&\, 0.51750&\, 0.17618&\, -0.009406&\, -0.009553\\
0.6&\, 0.49427&\, 0.16917&\, -0.007774&\, -0.007880\\
0.5&\, 0.47666&\, 0.16349&\, -0.006014&\, -0.006092
\end{tabular}
\end{center}
\caption{Variational parameters $\eta$, $\tau$ and correlation energy per site for $U=1$, $L=1000$ and various particle densities $n$,
 in comparison with the correlation energy $E_{\mbox{\scriptsize corr}}^{({\mbox{\scriptsize ex}})}$ obtained by second-order perturbation theory ($U=1$, $L=1000$).}
\label{table:normal}
\end{table}

It is instructive to compare these results with the exact second-order term deduced by perturbation theory \cite{Metzner_89}. The latter can be written as
\begin{align}
E^{(\mbox{\scriptsize ex})}_{\mbox{\scriptsize corr}}&=\frac{U^2}{L^2}\sum_{{\bf k}_1, {\bf k}_2,{\bf k}_3,{\bf k}_4}\delta({\bf k}_1-{\bf k}_2+{\bf k}_3-{\bf k}_4)\nonumber\\&\qquad\qquad\qquad
\frac{P_{{\bf k}_1}Q_{{\bf k}_2}P_{{\bf k}_3}Q_{{\bf k}_4}}{\xi_{{\bf k}_1}-\xi_{{\bf k}_2}+\xi_{{\bf k}_3}-\xi_{{\bf k}_4}}\nonumber\\
&=-N_sU^2\int_0^\infty d\tau\sum_{\bf R}[P({\bf R},\tau)Q({\bf R},\tau)]^2\, .
\label{eq:exact}
\end{align}
The data shown in Table \ref{table:normal} confirm that our variational ansatz reproduces the exact second-order term to a high precision ($97\%$ at half filling, $99\%$ for $n=0.5$). The corresponding results for the Gutzwiller ansatz, listed in Table \ref{table:gutz}, are substantially less accurate ($81\%$ of the exact correlation energy at half filling and 86\% for $n=0.5$). This large improvement of the correlation energy by the parameter $\tau$ also holds for larger values of $U$\, \cite{Otsuka_92, Dzierzawa_95}. We notice that the $\eta$ values for the variational ansatz including $\tau$ are much larger than those for the Gutzwiller ansatz (where $\tau=0$). This happens because with increasing $\tau$ the contribution of the region away from the Fermi surface is reduced and thus the cost in band energy due to the reduction of double occupancy is lowered, and $\eta$ assumes higher values than for $\tau=0$.

The relatively large values of the correlation parameter $\eta$ for small values of $U$ reflect the fact that small bare interactions do not necessarily imply weak correlations.
\begin{table}
\begin{center}
\begin{tabular}{llll}
$\, n$&\quad$\eta^{(G)}$& $\quad E_{\mbox{\scriptsize corr}}^{(G)}/L$&$\quad E_{\mbox{\scriptsize corr}}^{(\mbox{\scriptsize ex})}/L$\\
\hline
1.0&\, 0.14717  &\, -0.010220&\, -0.012562\\
0.9&\, 0.14639 &\, -0.009961&\, -0.012072\\
0.8&\, 0.14457  &\, -0.009236&\, -0.010995\\
0.7&\, 0.14216 &\, -0.008139&\, -0.009553\\
0.6&\, 0.13944 &\, -0.006771&\, -0.007880\\
0.5&\, 0.13652  &\, -0.005246&\, -0.006092
\end{tabular}
\end{center}
\caption{Variational parameter $\eta^{(G)}$ and correlation energy  $E_{\mbox{\scriptsize corr}}^{(G)}$ for the Gutzwiller ansatz, in comparison with the correlation energy $E_{\mbox{\scriptsize corr}}^{({\mbox{\scriptsize ex}})}$ obtained by second-order perturbation theory ($U=1$, $L=1000$).}
\label{table:gutz}
\end{table}

\section{Superconducting state}
\label{4}
\subsection{Reference state}
\label{41}
For superconductivity the reference state is the ground state of the mean-field Hamiltonian
\begin{align}
H_m=&\sum_{\bf k}\{\xi_{\bf k}(c_{\bf k\uparrow}^\dag c_{\bf k\uparrow}^{\phantom{}}
+c_{\bf -k\downarrow}^\dag c_{\bf -k\downarrow}^{\phantom{}})\nonumber\\
&\qquad-\Delta_{\bf k}(c_{\bf k\uparrow}^\dag c_{\bf -k\downarrow}^\dag
+c_{\bf -k\downarrow}^{\phantom{}}c_{\bf k\uparrow}^{\phantom{}})\}\, ,
\label{eq:ham_mf}
\end{align}
where $\xi_{\bf k}:=\varepsilon_{\bf k}-\mu$ and the gap parameter $\Delta_{\bf k}$ must have an appropriate symmetry, such as $d$-wave or $p$-wave. In this paper we restrict ourselves to $d$-wave symmetry, i.e., 
\begin{align}
\Delta_{\bf k}=\Delta_0(\cos k_x-\cos k_y)\, .
\label{eq:gap}
\end{align}
The parameter $\mu$ is chosen in such a way that the average particle number is equal to a fixed value. For the correlated ground state (\ref{eq:ansatz}) $\mu$ can be identified with the (true) chemical potential only for $\Delta_{\bf k}= 0$ and $U=0$.

$H_m$ is diagonalized by the Bogoliubov transformation
\begin{align}
c_{\bf k\uparrow}^{\phantom{}}&=\cos\vartheta_{\bf k}\, \alpha_{\bf k\uparrow}^{\phantom{}}
+\sin\vartheta_{\bf k}\, \alpha_{\bf -k\downarrow}^\dag\, ,\nonumber\\
c_{\bf -k\downarrow}^\dag&=-\sin\vartheta_{\bf k}\, \alpha_{\bf k\uparrow}^{\phantom{}}
+\cos\vartheta_{\bf k}\, \alpha_{\bf -k\downarrow}^\dag\, ,
\label{eq:bogoliubov}
\end{align}
if $\vartheta_{\bf k}$ is chosen as
\begin{align}
\cos 2\vartheta_{\bf k}=\frac{\xi_{\bf k}}{E_{\bf k}}\, ,\quad 
\sin 2\vartheta_{\bf k}=\frac{\Delta_{\bf k}}{E_{\bf k}}\, ,
\label{eq:theta}
\end{align}
where
\begin{align}
E_{\bf k}=\sqrt{\xi_{\bf k}^2+\Delta_{\bf k}^2}
\label{eq:spectrum}
\end{align}
is the excitation spectrum. The mean-field Hamiltonian now reads
\begin{align}
H_m=\sum_{\bf k}(\xi_{\bf k}-E_{\bf k})+\sum_{\bf k\sigma}E_{\bf k}\, 
\alpha_{\bf k\sigma}^\dag \alpha_{\bf k\sigma}\, .
\label{eq:ham_mf_diag}
\end{align}
Its ground state $\vert\Psi_m\rangle$ is the vacuum for quasi-particles, 
$\alpha_{{\bf k}\sigma}^{\phantom{}}\vert\Psi_m\rangle=0$. It is then easy to see that
\begin{align}
c_{\bf k\sigma}^{\phantom{}}(\tau)\vert\Psi_m\rangle=e^{-\tau E_k}c_{\bf k\sigma}^{\phantom{}}\vert\Psi_m\rangle,\nonumber\\
c_{\bf k\sigma}^\dag(\tau)\vert\Psi_m\rangle=e^{-\tau E_k}c_{\bf k\sigma}^\dag\vert\Psi_m\rangle\, .
\end{align}

Three different functions appear in the Wick decomposition, the momentum distribution functions
\begin{align}
P_{\bf k}:&=\langle c_{\bf k\sigma}^\dag c_{\bf k\sigma}^{\phantom{}}\rangle=\frac{E_{\bf k}-\xi_{\bf k}}{2E_{\bf k}}\, ,\nonumber\\
Q_{\bf k}:&=\langle c_{\bf k\sigma}^{\phantom{}} c_{\bf k\sigma}^\dag\rangle=\frac{E_{\bf k}+\xi_{\bf k}}{2E_{\bf k}}\, ,
\label{eq:pk}
\end{align}
and the ``Gor'kov function''
\begin{align}
F_{\bf k}:&=\langle c_{\bf -k\downarrow}^{\phantom{}} c_{\bf k\uparrow}^{\phantom{}}\rangle
=\langle c_{\bf k\uparrow}^\dag c_{\bf -k\downarrow}^\dag\rangle=\frac{\Delta_{\bf k}}{2E_{\bf k}}\, .
\label{eq:fk}
\end{align}
\subsection{Second-order expansion}
\label{42}
We are now prepared for carrying out explicitly the expansion (\ref{eq:expansion}) for a superconducting reference state. The contribution of zeroth order is given by
\begin{align}
\langle H\rangle=2\sum_{\bf k}\varepsilon_{\bf k}P_{\bf k}+N_sU\left(\frac{n^2}{4}+f_0^2\right)\, ,
\end{align}
where 
\begin{align}
f_0:=\frac{1}{N_s}\sum_{\bf k}F_{\bf k}\, .
\label{eq:f0}
\end{align}
For an order parameter with $p$- or $d$-wave symmetry, the ``average Gor'kov function'' $f_0$ vanishes if both the lattice and the boundary conditions have fourfold rotational symmetry. In order to cope with slight deviations from this symmetry for finite system sizes (due to periodic-antiperiodic boundary conditions) we retain $f_0$ in the analytical expressions. For the large system sizes considered here the breaking of the fourfold rotational symmetry has very little effect. In recent variational Monte Carlo studies \cite{Ido_18}, with periodic-antiperiodic boundary conditions and $L$ up to 24, striped phases have been found to be slightly more stable than homogeneous superconductivity. Because breaking of the fourfold rotational symmetry is expected to favor stripes and to weaken $d$-wave superconductivity, one may ask whether these results survive for larger system sizes or for symmetric boundary conditions.

The contribution of first order in $\eta$ has three terms
\begin{align}
\langle HD(\tau)\rangle_c=A+U(B+C)\, ,
\end{align}
where $A$ comes from the hopping part of the Hamiltonian and $B,C$ from the interaction. They correspond to the three diagrams of 
Fig. \ref{fig:first_order}, where a line can represent any of the three functions $P,Q,F$, and are given explicitly by 
\begin{align}
A&=2\sum_{\bf k}e^{-2\tau E_{\bf k}}\varepsilon_{\bf k}F_{\bf k}S_{\bf k}\, ,\nonumber\\
B&=\sum_{\bf k}e^{-2\tau E_{\bf k}}S_{\bf k}^2\, ,\nonumber\\
C&=\frac{1}{N_s^2}\sum_{\bf p,q,l,k}e^{-\tau(E_{\bf p}+E_{\bf q}+E_{\bf l}+E_{\bf k})}\nonumber\\
&\qquad\qquad\qquad\delta({\bf p+q+l+k})\nonumber\\ &\qquad(P_{\bf p}Q_{\bf q}-F_{\bf p}F_{\bf q})(P_{\bf l}Q_{\bf k}-F_{\bf l}F_{\bf k})\, ,
\label{eq:first}
\end{align}
where
\begin{align}
S_{\bf k}=n F_{\bf k}+f_0G_{\bf k}
\label{eq:sk}
\end{align}
with
\begin{align}
G_{\bf k}=Q_{\bf k}-P_{\bf k}=\frac{\xi_{\bf k}}{E_{\bf k}}\, .
\label{eq:gk}
\end{align}

The triple summation in $C$ is replaced by a simple summation over lattice sites using Eq. (\ref{eq:Kronecker}) together with the Fourier transform
\begin{align}
P({\bf R},\tau):=\frac{1}{N_s}\sum_{\bf k}e^{i{\bf k\cdot R}}e^{-\tau E_{\bf k}}P_{\bf k}
\end{align}
and correspondingly for the other functions. We get
\begin{align}
C=N_s\sum_{\bf R}\big[ P({\bf R},\tau)Q({\bf R},\tau)-F^2({\bf R},\tau)\big]^2\, .
\end{align}

We now turn to the second-order contribution in Eq. (\ref{eq:expansion}). The diagrams are grouped according to the three general structures of Fig. \ref{fig:second_order} but, because of the various possibilities for lines (representing $P_{\bf k}$, $Q_{\bf k}$ or $F_{\bf k}$) and Hartree bubbles (density per spin or average Gor'kov function) there are many different specific diagrams, namely 29 of type $A$, 18 of type $B$ and 16 of type $C$. Nevertheless, the result can be presented in a relatively compact form, as shown in Appendix \ref{A}. The numerical evaluation of the various terms requires only simple summations, either in $k$- or in $R$-space.

The second-order expansion of the particle number, Eq. (\ref{eq:expansionnumber}) is effectuated in the same way. To deduce the corresponding formulae one simply has to replace $\varepsilon_{\bf k}$ by 1 and $U$ by 0 in the expression for the energy.
\subsection{Numerical procedure}
\label{43}
In the numerical calculations we have considered finite quadratic arrays of size $L\times L$ with $L$ up to 4000. Thus the number  of ${\bf k}$-points in the Brillouin zone is $N_s=L^2$. Periodic-antiperiodic boundary conditions have been used, in order to reduce level degeneracies. 
For a given density $n=\overline{N}/N_s$ and a given system size the particle number $\overline{N}$ is chosen in such a way that there be no ambiguity in the reference state at $\Delta_0=0$. For $n=0.8$ and $L=1000$ this is the case for $\overline{N}=800'000$ because with this choice the ``highest occupied molecular orbital'' (HOMO) is completely full and the ``lowest unoccupied molecular orbital'' (LUMO) is completely empty and there is no degeneracy in the reference state. For a lattice of $100\times 100$ sites a particle number of 8000 would not lead to a full-shell situation, the closest numbers satisfying this criterion are $\overline{N}=7996$ and  $\overline{N}=8004$. For a finite gap parameter the constraint of a fixed particle number has to be satisfied very accurately, because the energy gained by pairing, the ``condensation energy'', is much smaller than the correlation energy. The results presented below have been obtained with a precision of at least $10^{-14}$ for the density $n$.

Four parameters have to be determined by minimizing the energy for a fixed density, namely $\tau$, $\eta$, $\Delta_0$ and $\mu$. To reduce the complexity of the problem, we use the fact that the parameters $\tau$ and $\Delta_0$ interfere weakly. Therefore we determine the optimal value of $\tau$ initially, i.e., for $\Delta_0=0$. We have checked that a full variational treatment of all parameters would only slightly increase the stability of the superconducting state. This is illustrated in Table \ref{table:optimal}, where the full optimization is shown to enhance the gap parameter by about $1\%$. Correspondingly, the condensation energy increases slightly. In what follows, we will restrict ourselves to the ``initial optimization''.  Some examples of optimized parameters are given in Table \ref{table:u} for $0.6\le U\le 1.2$. The parameter $\tau_0$ is practically $U$-independent in this range, while the correlation parameter $\eta$ is proportional to $U$ and  the gap parameter $\Delta_0$ varies much more strongly. A more detailed discussion of the $U$-dependence of the gap parameter will be given in Section \ref{6}.

\begin{table}
\centering
\begin{tabular}{lllll}
\, $n$&\quad $\tau_0$&$\quad \Delta_0(\tau_0)$&\quad $\tau_{\mbox{\scriptsize opt}}$&\, $\Delta_0( \tau_{\mbox{\scriptsize opt}})$\\
\hline
1.0&\, 0.198053&\, 0.0004247&\, 0.198882&\, 0.0004252\\
0.8&\, 0.184349&\, 0.0026698&\, 0.186988&\, 0.0027491\\
0.6&\, 0.169172&\, 0.0020630&\, 0.170660&\, 0.0020878
\end{tabular}
\caption{Gap parameters in the cases of full ($\tau_{\mbox{\scriptsize opt}}$) and ``initial'' optimization $(\tau_0)$, for $U=1$, $L=1000$ and three different densities $n$.}
\label{table:optimal}
\end{table}

\begin{table}
\centering
\begin{tabular}{llll}
$\, U$&$\quad \tau_0$&\quad $\eta$&\quad $\Delta_0$\\
\hline
0.6&\, 0.184349&\, 0.327026&\, 0.0004204\\
0.8&\, 0.184349&\, 0.436102&\, 0.0010891\\
1.0&\, 0.184349&\, 0.546204&\, 0.0026698\\
1.2&\, 0.184349&\, 0.660916&\, 0.0067672\\
\end{tabular}
\caption{Variational parameters $\tau_0, \eta, \Delta_0$ for $n=0.8$,  $L=1000$ and several values of $U$.}
\label{table:u}
\end{table}

\section{Energetics}
\label{5}
\subsection{Energy gain}
\label{51}
\begin{figure}
\centering
\includegraphics[width=7cm]{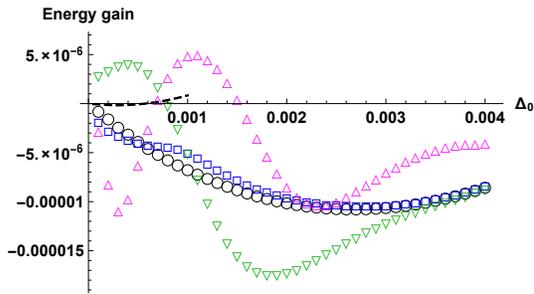}
\caption{Energy gain due to superconductivity as a function of the gap parameter $\Delta_0$ for $n=0.8$ and $U=1$. Symbols correspond to different system sizes, $L=1000$ (circles), $L=500$ (squares), $L=200$ (down-pointing triangles) and $L=100$ (up-pointing triangles). The dashed line represents results obtained with the Gutzwiller ansatz (for $L=1000$.)}
\label{fig:en}
\end{figure}

After the initial minimization with respect to $\tau$ for $\Delta_0=0$, the energy $\overline{E}(\eta,\mu,\Delta_0)$ is calculated for a fixed gap parameter 
$\Delta_0$ and minimized analytically with respect to $\eta$, while $\mu$ is determined by the constraint of a fixed density.
This yields the parameters $\eta(\Delta_0)$, $\mu(\Delta_0)$ and the energy difference
\begin{align}
\varepsilon(\Delta_0):=\frac{1}{N_s}[\overline{E}(\Delta_0)-\overline{E}(0)]\, . 
\end{align}
Fig. \ref{fig:en} shows this quantity for $U=1$, $n=0.8$ and four different system sizes. Negative values of $\varepsilon(\Delta_0)$ imply that the system is unstable with respect to superconductivity. Both for $L=100$ and for $L=200$, $\varepsilon(\Delta_0)$ exhibits two minima, but the two curves differ appreciably from each other and from curves for larger system sizes. By contrast, the results for $L=500$ and $L=1000$ are quite similar, with a single minimum at $\Delta_0\approx 0.0027$. These finite-size effects will be discussed in more detail in Section \ref{8}. 

For comparison, $\varepsilon(\Delta_0)$  is also presented in Fig. \ref{fig:en} for the Gutzwiller ansatz ($\tau=0$). A minimum is again found, but its position is an order of magnitude below that obtained for finite $\tau$. Correspondingly, the energy gain is nearly two orders of magnitude smaller. Therefore the variational parameter $\tau$ enhances both the correlation energy (as shown in Section \ref{2}) and the energy gain due to superconductivity. This is an important observation because one could imagine a poor correlation energy to be compensated by an artificially large condensation energy. 
\subsection{Condensation energy}
\label{52}
\begin{figure}
\centering
\includegraphics[width=7cm]{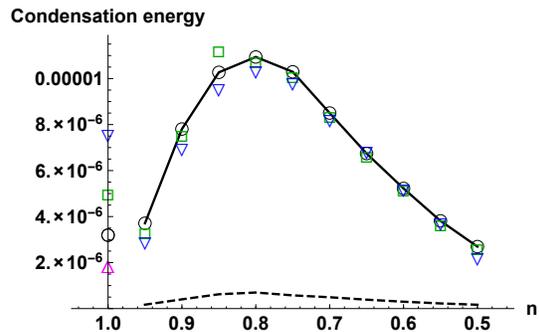}
\caption{Condensation energy for $U=1$ and different system sizes. Down-pointing triangles: $L=500$, squares: $L=1000$, circles and solid line: $L=2000$, up-pointing triangle: $L=4000$. The dashed line represents the BCS prediction, Eq. (\ref{eq:econdbcs}).}
\label{fig:econd}
\end{figure}

The minimization of the energy yields the optimized values of the parameters $\Delta_0, \eta$ and $\mu$. The optimized gap parameter $\Delta_0$ \footnote{From now on the symbol $\Delta_0$ will be used for the optimized gap parameter.} will be detailed in Section \ref{6}. Fig. \ref{fig:econd} shows the condensation energy, which we define as  
\begin{align}
\varepsilon_{\mbox{\scriptsize cond}}=\frac{1}{N_s}[\overline{E}(0)-\overline{E}_{\mbox{\scriptsize min}}(\Delta_0)]\, , 
\end{align}
as in BCS theory \cite{Bardeen_57, Schrieffer_64}.
$\varepsilon_{\mbox{\scriptsize cond}}$ first increases when moving away from half filling (i.e., upon doping the parent half-filled system), passes through a maximum for $n\approx 0.8$ and then decreases. The values differ little when passing from $L=500$ to $L=2000$, except at half filling where 
$\varepsilon_{\mbox{\scriptsize cond}}$ decreases with increasing $L$. The calculations have been stopped at $n=0.5$. One expects that the condensation energy would decrease further for smaller densities, but in this case one would have to take into account the competing superconducting ground state with $p$-wave symmetry.

In BCS theory the condensation energy is related to the density of states $N(0)$ and the gap parameter $\Delta_0$ 
by the simple formula
\begin{align}
\varepsilon_{\mbox{\scriptsize cond}}^{(\mbox{\scriptsize BCS})}=\frac{N(0)}{2}\Delta_0^2\, .
\label{eq:econdbcs}
\end{align}
Our results (solid line in Fig. \ref{fig:econd}) differ markedly from the BCS prediction (dashed line in Fig. \ref{fig:econd}). To find out whether the discrepancy can be attributed to the symmetry of the gap function ($s$-wave due to a local attraction in BCS theory, $d$-wave in the present case), we have calculated the condensation energy for an extended Hubbard model with nearest-neighbor attraction using the mean-field ground state of section \ref{41}. Details are given in Appendix \ref{B}. The results agree quite well with Eq. (\ref{eq:econdbcs}), which seems to hold approximately for any BCS-type theory. But why is the condensation energy so much larger in the present case than what would be predicted by BCS theory? We attribute the difference to the correlation energy, which involves not only the region very close to the Fermi energy, but also band states further away. In fact, the correlation parameter $\eta$ - and therefore the correlation energy - increases with $\Delta_0$, as can be verified by comparing Tables \ref{table:normal} and \ref{table:u} (for $U=1$). The disparity would be even more pronounced if the fully optimized value of $\tau$ would be used for the superconducting state.

\subsection{Conventional or unconventional?}
\label{53}

Unconventional superconductivity is not a sharply defined concept. Sometimes it is associated with an unconventional symmetry of the order parameter, and sometimes the emphasis is on properties deviating markedly from BCS predictions or on the non-phonon glue mediating the effective attraction \cite{Sigrist_91, Stewart_17}. Superconductivity in the repulsive Hubbard model is unconventional in several respects, in its order parameter ($d$-wave close to half filling, $p$-wave further away), in the mechanism (no phonons by assumption, maybe exchange of spin fluctuations or no glue at all) and also in deviations from BCS predictions. Here we discuss the question whether pairing is due to a decrease in potential energy, as in BCS theory, or rather due to an unconventional decrease in kinetic energy.

For the reduced BCS Hamiltonian it is easy to convince oneself that the kinetic energy cannot be lowered by pairing. For this model the normal state corresponds to the filled Fermi sea, which has the lowest possible kinetic energy for a given number of particles. Any interaction effect must then lead to an increase of kinetic energy, and superconductivity can only occur if this increase is overcompensated by a decrease in potential energy.

The issue whether the condensation energy is generated by a gain in potential energy, as in BCS theory, or by a gain in kinetic energy has been addressed in the frameworks of spin-fluctuation exchange \cite{Yanase_05}, cluster dynamical mean-field theory
\cite{Maier_04, Gull_12} and variational methods \cite{Yokoyama_04, Yokoyama_13, Kaczmarczyk_13, Tocchio_16}. Quite generally, an unconventional gain in kinetic energy is found for large values of $U$ and/or weak doping, while a conventional gain in potential energy is obtained for heavy doping and/or not too large $U$, but the detailed predictions differ somewhat. For instance, different variational wave functions may give different answers for the same values of $U$ and for the same density $n$\, \cite{Eichenberger_07}.

\begin{figure}
\centering
\includegraphics[width=7cm]{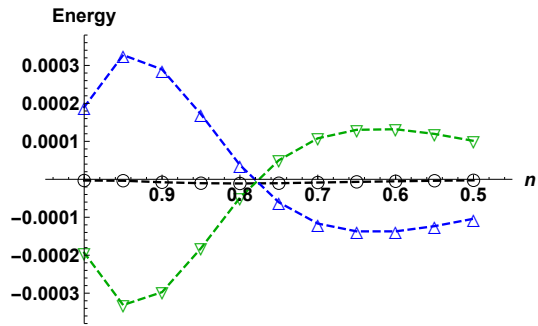}
\caption{Changes in kinetic (up-pointing triangles), potential (down-pointing triangles) and total energies (circles) for $U=1$ and $L=2000$.}
\label{fig:kinpot}
\end{figure}

It is straightforward to calculate individually the changes in potential and kinetic energies due to superconductivity within the present approach. The results, shown in Fig. \ref{fig:kinpot} for $U=1$, are quite surprising, because the kinetic energy is lowered for heavy doping ($n\lesssim 0.78$), while for weak doping ($n\gtrsim 0.78$) there is a gain in potential energy, contrary to what is typically found in the numerical calculations mentioned above. There is no contradiction with the arguments given for the reduced BCS Hamiltonian, because the normal state ($\Delta_0=0$) is correlated and has a kinetic energy exceeding its minimum value. Fig. \ref{fig:kinpot} also shows that the individual changes in potential and kinetic energies are much larger than the condensation energy. The same delicate balance between kinetic and potential energies has been observed some time ago on the basis of both variational Monte Carlo \cite{Ogata_06} and dynamic cluster calculations \cite{Gull_12}.
The corresponding results for the Gutzwiller ansatz will be presented in Section \ref{9} and shown to be quite similar. 

\begin{figure}
\centering
\includegraphics[width=7cm]{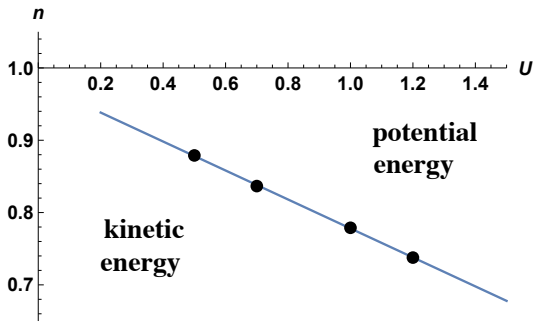}
\caption{Crossover from unconventional (``kinetic-energy lowering'') to conventional (``potential-energy lowering'') superconductivity for very small values of $U$. The numerical results (dots) have been calculated for $L_x=2000$ and are well converged. The line is a linear interpolation of the data points.}
\label{fig:kinpot2}
\end{figure}

In order to scrutinize the unexpected kinetic-energy lowering, we have determined the crossing point between conventional and unconventional regimes for several values of $U$. The results shown in Fig. \ref{fig:kinpot2} indicate that for small values of $U$ kinetic-energy lowering is found for densities sufficiently far away from half filling. This is a region where umklapp scattering is expected to play a minor role, and we may speculate that this is the reason why electrons are ``happy moving together''  \cite{Hirsch_02}.
Fig. \ref{fig:kinpot2} also shows that there is no real discrepancy with reports of conventional behavior for small values of $U$, because this conclusion is commonly reached on the basis of results obtained for $U\gtrsim 4$\, \cite{Kaczmarczyk_13, Yokoyama_04, Yokoyama_13, Yanase_05, Maier_04, Gull_12, Tocchio_16, Fratino_16}.

\section{Gap parameter}
\label{6}
\begin{figure}
\centering
\includegraphics[width=7cm]{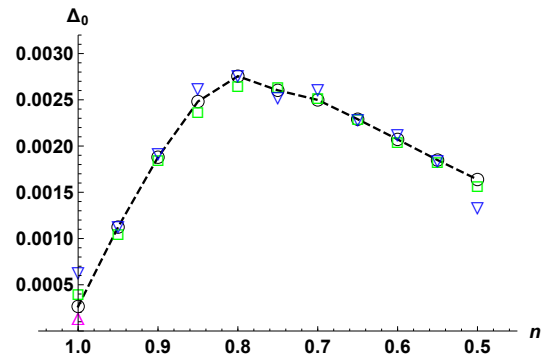}
\caption{Gap parameter as a function of electron density for $U=1$ and different system sizes. Down-pointing triangles: $L=500$, squares: $L=1000$, circles and dashed line: $L=2000$, up-pointing triangle: $L=4000$.}
\label{fig:gap}
\end{figure}

The optimized gap parameter $\Delta_0$ is given in Fig. \ref{fig:gap} for $U=1$ as a function of density for various system sizes. It shows a similar dome-like shape as the condensation energy (depicted in Fig. \ref{fig:econd}), although less pronounced. Again the results vary little with system size from $L=500$ to $L=2000$ except at half filling where they decrease steadily. Finite-size scaling (discussed in Section \ref{8}) indicates that at half filling the gap parameter vanishes in the thermodynamic limit. 

An interesting question is how $\Delta_0$ varies with $U$. Adopting the RPA expression for the effective interaction induced by the exchange of spin fluctuations \cite{Scalapino_95, Kondo_01} in the small $U$ limit, one obtains an attraction proportional to $U^2$ and a BCS behavior for the gap parameter,
\begin{align}
\log \frac{1}{\Delta_0}\propto \frac{1}{U^2}\, .
\label{eq:eliashberg}
\end{align}
The same dependence on $U$ has been found for the critical temperature (which is expected to be proportional to $\Delta_0$) using a renormalization group approach \cite{Raghu_10}.
 
\begin{figure}
\centering
\includegraphics[width=7cm]{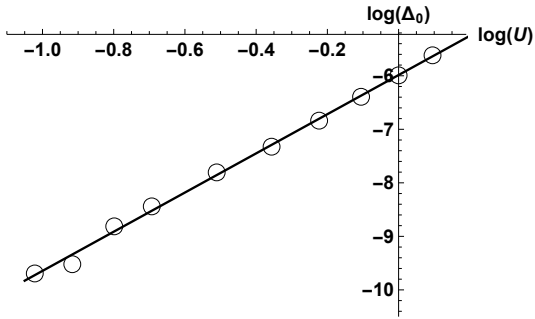}
\caption{Gap parameter as a function of $U$ for $n=0.7$ and $L=2000$. The solid line is a linear fit through the data points.} 
\label{fig:gap(U)}
\end{figure}

To see whether the $U$-dependence of Eq. (\ref{eq:eliashberg}) also comes out from our variational method, we have calculated the gap parameter for various values of $U$. These are limited to a relatively small region because for too large values of $U$ the second-order expansion breaks down and for too small values of $U$ the tiny condensation energy would require a higher precision than what we used in the calculations. Fig. \ref{fig:gap(U)} shows results for $n=0.7$ and $0.36\le U\le 1.1$. A linear relationship between 
$\log \Delta_0$ and $\log U$ fits the data very well, giving
\begin{align}
\Delta_0=(U/U^*)^\gamma\, ,
\label{eq:powerlaw}
\end{align}
with $U^*\approx 5.2$ and $\gamma\approx 3.6$. By contrast, the functional behavior (\ref{eq:eliashberg}) does not fit these data. Results for $0.7<n<1$ are very similar, but because of the larger values of the correlation parameter $\eta$ (as compared to those for $n=0.7$) the range of
$U$-values has to be reduced. It is interesting to note that $U^*$ is of the order of the interaction strength where a Mott transition would occur in the absence of antiferromagnetic ordering \cite{Fratino_17}. We notice that a power law has also been extracted from variational Monte Carlo calculations \cite{Yokoyama_04, Ogata_06}.

The apparent discrepancy between Eqs. (\ref{eq:eliashberg}) and (\ref{eq:powerlaw}) can have various causes. First, the values of $U$ used in the present calculations could still lie outside of the asymptotic region, and therefore a crossover to exponential behavior at still smaller values of $U$ cannot be excluded. Second, the ``superconducting gap'' has quite different meanings in different approaches. Thus in the renormalization group approach of Ref. \onlinecite{Raghu_10} the gap is assumed to have the same asymptotic behavior as the characteristic energy scale  below which the approach breaks down, while in the present approach $\Delta_0$ is a variational parameter. The two methods could yield quite different gap parameters, which could also be quite different from the location of the peak in the spectral density. Third,  long-wavelength fluctuations have been neglected in the present approach. These could modify the $U$-dependence, in a similar way as already found for the antiferromagnetic order at half filling \cite{Schauerte_02}. Fourth, the functional renormalization group procedure cannot be worked out without approximations. The version used by Raghu and coworkers \cite{Raghu_10} to establish Eq. (\ref{eq:eliashberg}) works on a ``one-loop'' level and neglects both frequency-dependences and self-energy insertions in the irreducible vertices \cite{Shankar_94}. These approximations are not innocent \cite{Kugler_18}, but so far applications of more sophisticated ``multi-loop'' approaches have been limited to temperature-dependent generalized susceptibilities 
\cite{Tagliavini_19}.
Finally, we may wonder whether perturbative methods can yield the correct asymptotic behavior for 
$U\rightarrow 0$. We have already seen that the refinement of the Gutzwiller ansatz reduces dramatically the $U$-region where a low-order expansion in powers of the correlation parameter $\eta$ can be trusted. Further refinements towards the exact ground state could reduce further the convergence radius, which might vanish eventually. 
\section{Order parameter}
\label{7}
The physical interpretation of the gap parameter $\Delta_0$ is far from obvious for any variational treatment of a superconductor with strong correlations. Thus it cannot simply be associated with a pseudogap, although this may yield an appealing picture of weakly doped cuprates
\cite{Anderson_04}. In fact, the value of $\Delta_0$ depends quite strongly on the choice of the wave function \cite{Eichenberger_07}.
By contrast, the order parameter can be sharply defined as an expectation value, and it depends much less on details \cite{Baeriswyl_09}. In ``canonical'' calculations (fixed particle number) the order parameter is deduced from the long-distance behavior of the pair-pair correlation function \cite{Vondelft_01}. In the present ``grand-canonical'' approach it is defined as the pair amplitude on nearest-neighbor sites ${\bf R, R'}$,
\begin{align}
\Phi:=\frac{\langle\Psi\vert c_{{\bf R}\uparrow}^\dag c_{{\bf R'}\downarrow}^\dag\vert\Psi\rangle}
{\langle\Psi\vert\Psi\rangle}
=\frac{1}{N_s}\sum_{\bf k}\cos k_x \frac{\langle\Psi\vert C_{\bf k}^\dag\vert\Psi\rangle}{\langle\Psi\vert\Psi\rangle}\, ,
\end{align}
where 
\begin{align}
C_{\bf k}:=c_{{\bf k}\uparrow}^\dag c_{-{\bf k}\downarrow}^\dag
\end{align}
creates a Cooper pair. The expansion in powers of $\eta$ proceeds in exactly the same way as for the hopping term and we find
\begin{align}
&\frac{\langle\Psi\vert C_{\bf k}^\dag\vert\Psi\rangle}{\langle\Psi\vert\Psi\rangle}=F_{\bf k}
-\eta\langle(C_{\bf k}^\dag+C_{\bf k}^{\phantom{}})D(\tau)\rangle_c
\nonumber\\
&\qquad +\frac{\eta^2}{2}\big[\langle(C_{\bf k}^\dag+C_{\bf k}^{\phantom{}})D^2(\tau)\rangle_c\nonumber\\
&\qquad\quad +\langle D(-\tau)(C_{\bf k}^\dag+C_{\bf k}^{\phantom{}})D(\tau)\rangle_c\big]\, .
\label{eq:expop}
\end{align}

The zeroth-order term is just given by the Gor'kov function, as in BCS theory. The first-order term reads
\begin{align}
\langle(C_{\bf k}^\dag+C_{\bf k}^{\phantom{}})D(\tau)\rangle_c=e^{-2\tau E_{\bf k}}G_{\bf k}(n F_{\bf k}+f_0G_{\bf k})
\end{align}
and is represented by diagram $A$ of Fig. \ref{fig:first_order}. The second-order contributions correspond to the diagrams of Fig. \ref{fig:second_order}
and are given explicitly in Appendix \ref{C}. 
\begin{figure}
\centering
\includegraphics[width=7cm]{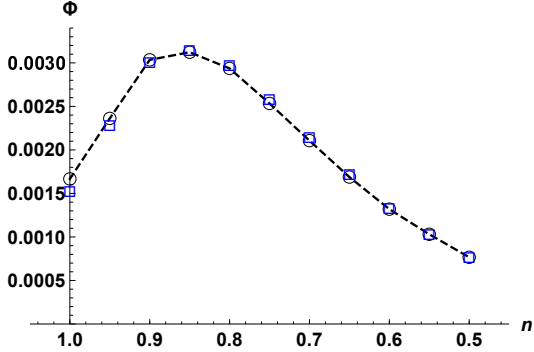}
\caption{Order parameter for $U=1$ and $L=1000$ as a function of the density $n$. Circles and the dashed line represent the second-order expansion, squares the zeroth-order contribution.}
\label{fig:op}
\end{figure}
The numerical evaluation of the order parameter $\Phi$ is again straightforward, as only simple $k$- and $R$-sums have to be calculated. The result for $U=1$ and $L=1000$ is shown in Fig. \ref{fig:op}. $\Phi$ has a maximum at $n\approx 0.85$ and is to a large extent given by the zeroth-order contribution (the Gor'kov function). For $n=1$ additional results for larger system sizes agree with the asymptotic behavior found for the gap parameter, which will be discussed below.

\begin{figure}
\centering
\includegraphics[width=7cm]{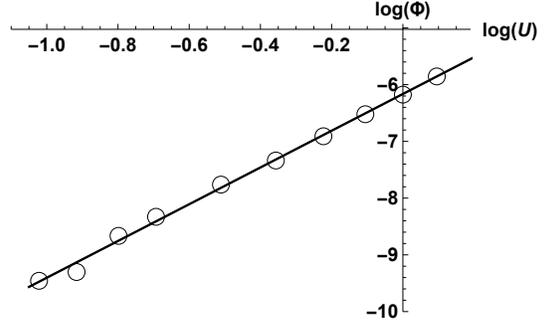}
\caption{Order parameter as a function of $U$ for $n=0.7$ and $L=2000$. The solid line is a linear fit through the data points.}
\label{fig:op2}
\end{figure}

Fig. \ref{fig:op2} shows the $U$-dependence of the order parameter for $n=0.7$ and $L=2000$. The behavior is nearly indistinguishable from that of the gap parameter (Fig. \ref{fig:op}) and again well described by the power law of Eq. (\ref{eq:powerlaw}), with $U^*\approx 6.7$ and 
$\gamma\approx 3.2$.

\section{Finite-size effects}
\label{8}

\begin{figure}
\centering
\includegraphics[width=7cm]{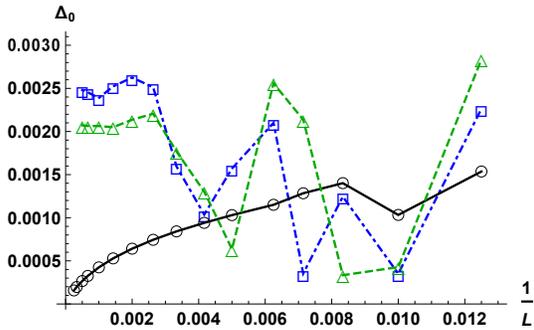}
\caption{Variation of the gap parameter with system size ($U=1$). Circles and solid line: $n=1$, squares: $n=0.85$, triangles: $n=0.6$.}
\label{fig:scaling}
\end{figure}

To highlight the size dependence of the gap parameter, we have plotted $\Delta_0$ vs. $L^{-1}$ for three different densities in Fig. \ref{fig:scaling}. Both for $n=0.85$ and  for $n=0.6$, $\Delta_0$ shows a rather irregular behavior for $L\lesssim 300$, but then it approaches a finite limiting value for the largest system sizes. The wild variation of $\Delta_0$ as a function of $L$ for small to intermediate system sizes originates most likely from irregular changes of the HOMO-LUMO gap $\Delta_{\mbox{\scriptsize HL}}$ (the separation between the lowest unoccupied and the highest occupied single-particle levels $\varepsilon_{\bf k}$). 

\begin{figure}
\centering
\includegraphics[width=6cm]{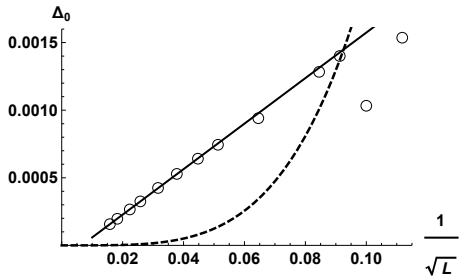}
\caption{Size dependence of the gap parameter for $U=1$ at half filling ($n=1$). The dashed line corresponds to the HOMO-LUMO gap 
$\Delta_{\mbox{\scriptsize HL}}$, given by Eq. (\ref{eq:HL}). The solid line is a linear fit through those data points for which $\Delta_0$ exceeds
$\Delta_{\mbox{\scriptsize HL}}$.}
\label{fig:halffilling}
\end{figure}

For $n=1$ the HOMO-LUMO gap is a smooth function,
\begin{align}
\Delta_{\mbox{\scriptsize HL}}=4\big(1-\cos\frac{\pi}{L}\big)\approx\frac{2\pi^2}{N_s}\, .
\label{eq:HL}
\end{align}
Correspondingly, as one can see from Fig. \ref{fig:scaling}, the superconducting gap $\Delta_0$ is also smooth as soon as the system size is large enough. The behavior for large $L$ is well described by the size dependence $\Delta_0\propto 1/\sqrt{L}$, as evidenced in Fig. \ref{fig:halffilling}, and therefore $\Delta_0$ vanishes for $L\rightarrow\infty$.
The figure also indicates that the boundary between regular and irregular behavior is defined by the equality of the two gaps, 
$\Delta_0=\Delta_{\mbox{\scriptsize HL}}$.

It is quite remarkable that superconductivity fades away at half filling, because we did not include the possibility of a competing antiferromagnetic instability in the variational ansatz. Had we done so, antiferromagnetic long-range order would readily show up \cite{Menteshashvili_14}, in agreement with conventional wisdom. The present results show that this competition is not necessary for quenching superconductivity at half filling. 

Already in 1959 Anderson wondered what happens to a superconducting material if its size shrinks more and more \cite{Anderson_59b}. He argued that superconductivity ceases as soon as the characteristic level spacing becomes larger than the energy gap of the bulk system. In the nineties, Anderson's question was investigated thoroughly, both in spectroscopic experiments on ultrasmall aluminium particles and theoretically using the exact solution of the reduced BCS Hamiltonian \cite{Vondelft_01}. Anderson's estimate was confirmed, at the same time the critical size was found to signal a crossover rather than a true transition.

\begin{figure}
\centering
\includegraphics[width=7cm]{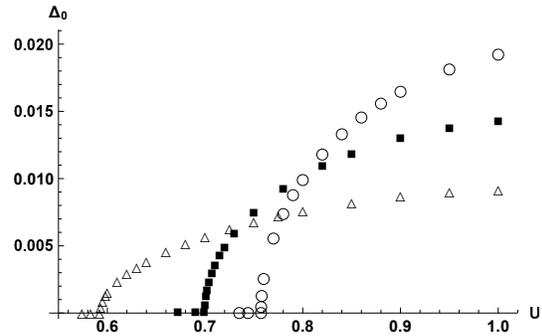}
\caption{$U$-dependence of the gap parameter for $n=0.85$ and small system sizes, $L=12$ (circles), $L=14$ (squares) and $L=20$ (triangles).}
\label{fig:uc1}
\end{figure}

\begin{figure}
\centering
\includegraphics[width=7cm]{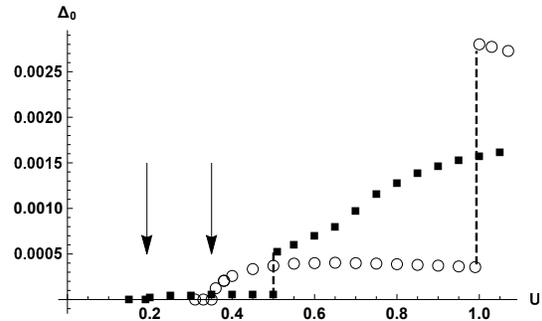}
\caption{$U$-dependence of the gap parameter for $n=0.85$ and intermediate system sizes, $L=100$ (circles) and $L=200$ (squares). The arrows indicate the critical points above which $\Delta_0$ is finite.}
\label{fig:uc2}
\end{figure}

\begin{figure}
\centering
\includegraphics[width=7cm]{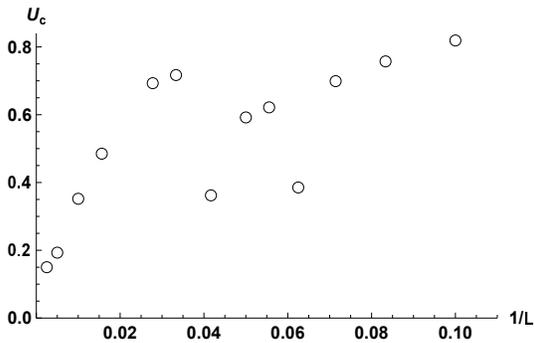}
\caption{Size dependence of the critical value $U_c$, below which superconducting order disappears, for $n=0.85$.}
\label{fig:uc}
\end{figure}

We now address Anderson's question in the present context and search for the minimal length above which our wave function has superconducting order. Because of the irregularities described above (away from half filling) we do not vary $L$ for a fixed value of $U$, we rather vary $U$ for a fixed value of $L$. Therefore we determine the critical value of $U$, above which $\Delta_0$ is finite. Fig. \ref{fig:uc1} shows the results for three small systems and a density of 0.85. The gap parameter grows continuously from zero above a critical value $U_c(L)$. The larger $L$ the smaller $U_c(L)$. As to the numerical values of $U_c$, they are nearly an order of magnitude smaller than predicted by Anderson's criterion. Fig. \ref{fig:uc2} shows similar results for systems of intermediate size, $L=100$ and $200$. While $\Delta_0$ again increases first smoothly, there exist discontinuities, which result from energy curves with two minima, such as that shown in Fig. \ref{fig:en}. On the left side of a jump the lower minimum has lower energy, on the right side the upper minimum is more stable. On average, the critical values $U_c$ decrease with system size, as exemplified in Fig. \ref{fig:uc} for $n=0.85$. The data shown in the figure are consistent with a vanishing $U_c$ for $L\rightarrow\infty$, i.e., for an infinite system there is superconductivity for arbitrarily small $U$.

\begin{figure}
\centering
\includegraphics[width=7cm]{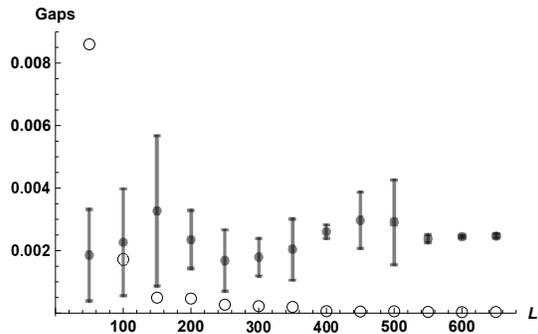}
\caption{Average gap parameter $\Delta_0$ (dots) and standard deviation (error bars) for $n=0.85$, $U=1$ and various (average) system sizes. The circles indicate the average HOMO-LUMO gap  $\Delta_{\mbox{\scriptsize HL}}$. Its standard deviation is of the order of the average value. }
\label{fig:cg}
\end{figure}

To explore the fluctuations of the gap parameter as a function of system size in a more quantitative way than above, we introduce coarse-graining for both gaps ($\Delta_0$ and $\Delta_{\mbox{\scriptsize HL}}$), by averaging over five neighboring (even) values of $L$. Results for $n=0.85$ and $U=1$, shown in Fig. \ref{fig:cg}, reveal that the fluctuations remain large above Anderson's critical size ($L\approx 100$ in this case) and die out only when the superconducting gap spans several level spacings. Calculations for other values of $n$ and $U$ give similar results.

The overall evolution of superconducting order as a function of system size is thus  governed by two characteristic sizes, a first one where order appears continuously, much below Anderson's critical size, and a second one where the order parameter becomes well defined, far above Anderson's critical size. Similar finite-size effects should play a role in
any numerical treatment of the two-dimensional Hubbard model on a finite lattice, be it quantum Monte Carlo, variational treatments or dynamical mean-field theory. For $U$ of the order of the bandwidth ($U\approx 8$) the typical size of the superconducting gap is found to be 0.1\,  \cite{Eichenberger_07}. This is also the typical size of the HOMO-LUMO gap for a $12\times 12$ lattice. In this parameter regime fluctuations are expected to be substantial.

\section{Gutzwiller ansatz}
\label{9}
It is worthwhile to compare the results obtained for the variational ansatz (\ref{eq:ansatz}) with those deduced from the standard Gutzwiller wave function. The Gutzwiller predictions for energy and particle number can be found by putting $\tau=0$ in the corresponding expressions of Section \ref{4} and Appendix \ref{A}.
Because in this case the correlation parameter $\eta$ is much smaller than for the full ansatz, the expansion in powers of $\eta$ can be used for larger values of $U$. 

\begin{figure}
\centering
\includegraphics[width=7cm]{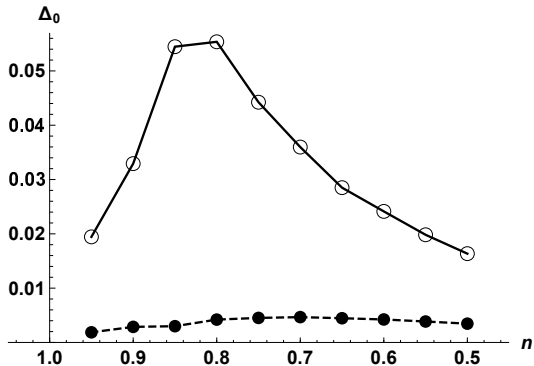}
\caption{Gap parameter as a function of electron density for $L=1000$, calculated with the Gutzwiller ansatz. Circles and solid line: $U=3$, dots and dashed line: $U=2$.}
\label{fig:gapGutz}
\end{figure}

Fig. \ref{fig:gapGutz} shows the optimized gap value as a function of density for $U=2$ and $U=3$, where $\eta$ is of the order of 0.3 and 0.45, respectively. The shape for $U=3$ is slightly more peaked than in the case of the full ansatz, Fig. \ref{fig:gap}, but the maximum occurs at about the same density. The gap values obtained for $U=2$ with the Gutzwiller ansatz are smaller than those of the full ansatz, which were calculated for $U=1$. This simply reflects the fact that superconductivity is strengthened by the additional term $e^{-\tau H_m}$ in Eq. (\ref{eq:ansatz}).

\begin{figure}
\centering
\includegraphics[width=7cm]{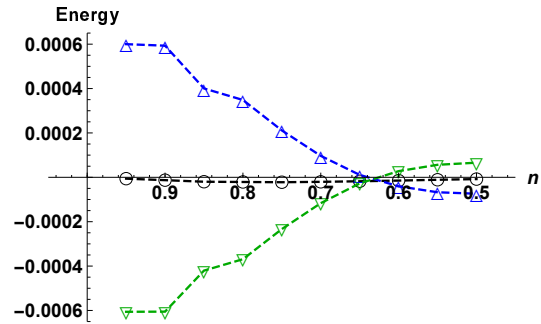}
\caption{Pairing-induced changes in kinetic (up-pointing triangles), potential (down-pointing triangles) and total energies (circles) for $U=2$ and $L=1000$, obtained with the Gutzwiller ansatz.}
\label{fig:kinpotGutz}
\end{figure}

Fig. \ref{fig:kinpotGutz} shows the condensation energy together with the changes of kinetic and potential energy due to pairing for $U=2$. One notices a striking similarity to Fig. \ref{fig:kinpot}. There is again a change from a potential- to kinetic-energy driven pairing as the density decreases, although now the crossing occurs at a lower density. For $U=3$ the behavior is conventional, i.e., 
there is a gain in potential energy for all densities between 1 and 0.5, accompanied by a loss in kinetic energy. 

Finite-size effects for the Gutzwiller wave function are quite similar to those described for the full ansatz in Section \ref{8}. At half filling, 
 we find again $\Delta_0\propto 1\sqrt{L}$ for large $L$. For other densities, there exists again a critical value $U_c$, below which $\Delta_0$ vanishes. For $U$ slightly above $U_c$ the behavior is like in Figs. \ref{fig:uc1} and \ref{fig:uc2}. However, the values for the onset of superconductivity are appreciably higher, as shown in Table \ref{table:onset}. Irregular behavior is also detected for the conventional Gutzwiller ansatz, although maybe slightly less violent than in Fig. \ref{fig:scaling}.

\begin{table}
\begin{tabular}{rll}
\, $L$&\, $U_c(\tau>0)$&$U_c(\tau=0)$\\
\hline
12& \quad 0.757&\quad 1.89\\
14&\quad 0.699&\quad 1.75\\
20&\quad 0.592&\quad 1.45\\
100&\quad 0.35&\quad 0.858\\
200&\quad 0.193&\quad 0.449\\
\end{tabular}
\caption{Critical values $U_c$ for $n=0.85$ and different system sizes. The values for the refined wave function (second column) are much smaller than those for the Gutzwiller ansatz (third column).}
\label{table:onset}
\end{table}

\section{Summary and outlook}
\label{10}
In this paper the $2D$ repulsive Hubbard model has been scrutinized for $d$-wave superconductivity, using a refined Gutzwiller ansatz for the ground state. The operator $e^{-\eta D}$, which simply reduces double occupancy of a reference state (a BCS mean-field state in the present case), was supplemented by a term 
$e^{-\tau H_m}$, which involves the BCS mean-field Hamiltonian $H_m$. This ansatz is well suited for treating the small $U$ region. On the one hand, it admits a linked-cluster expansion in powers of the correlation parameter $\eta$, as in the case of the standard Gutzwiller ansatz. On the other hand, the energy is greatly improved by the additional term, as evidenced by a comparison with the expansion of the exact ground-state energy. 
Moreover, this approach has the advantage that large system sizes (millions of sites) can be treated to a very high precision. The drawback is that the method is limited to a relatively small interval of coupling strengths, $U_1<U<U_2$, where $U_1\approx 0.35$ (for lattices which are not larger than a few million sites) and $U_2\approx 1.2$ (to keep $\eta$ smaller than about 0.6).  

The following main results were obtained: 
\begin{enumerate}
\item Superconductivity with $d$-wave symmetry exists away from half filling ($0.5\le n<1$) with a maximum stability for $n\approx 0.8$. 
\item Both the gap parameter $\Delta_0$ and the order parameter $\Phi$ are an order of magnitude larger than in the case of the Gutzwiller ansatz. \item Correspondingly, the condensation energy also has a maximum for $n\approx 0.8$ and is larger by nearly two orders of magnitude than in the Gutzwiller ansatz. 
\item The pairing is due to a lowering of the kinetic energy for small enough $U$ and for densities not too close to half filling.
\item For a given lattice size a minimum value $U_c$ is required for superconductivity to emerge. This value decreases steadily with increasing size. We conclude that for an infinite system superconductivity exists for arbitraliy small $U$.
\item Our findings resolve a discrepancy between variational Monte Carlo studies, which did not find signatures of superconductivity below some critical value of $U$ (of the order of 4) and perturbative treatments such as the functional renormalization group or the fluctuation-exchange approximation, which do find a superconducting instability for small values of $U$. The main reason for the failure of variational Monte Carlo calculations is that the system sizes that can be treated are not large enough for the small gap parameters found for $U\approx 1$.
\item In the special case of a half-filled band ($n=1$) $\Delta_0$ and $\Phi$ vanish as the system size tends to infinity. 
\end{enumerate}

Four variational parameters have been inserted into the trial wave function, the gap parameter $\Delta_0$, the ``chemical potential'' $\mu$, the correlation parameter $\eta$, and $\tau$, the inverse of a soft energy cut-off. The parameter $\tau$ can also be interpreted as a characteristic imaginary time. The exact second-order contribution, Eq. (\ref{eq:exact}), is indeed an integral over the imaginary time $\tau$. The corresponding variational term, Eq. (\ref{eq:1st}), looks very similar, but the integral is replaced by the integrand at ``time'' $\tau$. Since this characteristic time strengthens superconductivity, one may wonder whether its role is to introduce retardation in some effective interaction. It would be worthwhile to study this question thoroughly.

Other questions could also be addressed, such as $p$-wave superconductivity (which is expected to dominate for small densities $n$), the competition between superconductivity and antiferromagnetism (which is expected to be weak in view of the vanishing superconducting order parameter at half filling, where antiferromagnetic ordering is strongest) or the effect of next-nearest-neighbor hopping $t'$ (which could bring back superconductivity at half filling). 

\section*{Acknowledgment}
This work has been supported by the Swiss National Science Foundation Grant No. 200020-132698. I have profited from a close collaboration with Mikheil Menteshashvili at an early stage of this work and also from useful discussions with Florian Gebhard and J\"org B\"unemann. I also thank Hisatoshi Yokoyama and Masao Ogata for having provided their variational Monte Carlo data.
\begin{appendix}
\section{Second-order terms for the energy}
\label{A}
The second-order contributions of Eq. (\ref{eq:expansion}) can be grouped according to the diagrams of Fig. \ref{fig:second_order}, and
we write
\begin{align}
\langle H_0D^2(\tau)\rangle_c=A_1+B_1+C_1\, ,\nonumber\\
\langle D(-\tau)H_0D(\tau)\rangle_c=A_2+B_2+C_2\, .
\end{align}
Proceeding as in Section \ref{3}, we introduce the quantities
\begin{align}
\varepsilon_\nu({\bf R},2\tau)&=\frac{1}{N_s}\sum_{\bf k}e^{i{\bf k\cdot R}}\varepsilon_{\bf k}e^{-2\tau E_{\bf k}}g_{{\bf k}\nu}\, ,
\end{align}
where
\begin{align}
g_{{\bf k}\nu}=\left\{\begin{array}{ll}P_{\bf k}^2,&\nu=1\\
Q_{\bf k}^2,&\nu=2\\
F_{\bf k}^2,&\nu=3\\
F_{\bf k}G_{\bf k},&\nu=4\\
\end{array}\right.
\end{align}
All three diagrams of Fig. \ref{fig:second_order} contribute if the gap parameter is finite and we write, correspondingly,
\begin{align}
\langle H_0D^2(\tau)\rangle_c=A_1+B_1+C_1\, ,\nonumber\\
\langle D(-\tau)H_0D(\tau)\rangle_c=A_2+B_2+C_2\, .
\end{align}
We find
\begin{align}
A_1&=2\sum_{\bf k}e^{-2\tau E_{\bf k}}\varepsilon_{\bf k}F_{\bf k}S_{\bf k}(n G_{\bf k}-4f_0F_{\bf k})\, ,\nonumber\\
A_2&=2\sum_{\bf k}e^{-4\tau E_{\bf k}}\varepsilon_{\bf k}G_{\bf k}S_{\bf k}^2\, ,\nonumber\\
B_1&=\frac{2}{N_s}\sum_{\bf k,q}e^{-2\tau E_{\bf k}}\varepsilon_{\bf k}F_{\bf k}S_{\bf q}\nonumber\\
&\qquad\qquad\qquad(G_{\bf k}G_{\bf q}+4F_{\bf k}F_{\bf q})\, ,
\nonumber\\ 
B_2&=\frac{4}{N_s}\sum_{\bf k,q}e^{-2\tau (E_{\bf k}+2E_{\bf q})}\varepsilon_{\bf k}F_{\bf k}S_{\bf q}\nonumber\\
&\qquad\qquad(-Q_{\bf k}P_{\bf q}-P_{\bf k}Q_{\bf q}+2F_{\bf k}F_{\bf q})\, ,\nonumber\\ 
C_1&=4N_s\sum_{\bf R}\big[P({\bf R})Q({\bf R})-F^2({\bf R})\big]\nonumber\\
&\qquad\big[\varepsilon_3({\bf R},2\tau)G({\bf R})-\varepsilon_4({\bf R},2\tau)F({\bf R})\big]\, ,\nonumber\\
C_2&=2N_s\sum_{\bf R}\big[P({\bf R},2\tau)Q({\bf R},2\tau)-F^2({\bf R},2\tau)\big]\nonumber\\
\big[\varepsilon_2&({\bf R},2\tau)P({\bf R},2\tau)-\varepsilon_1({\bf R},2\tau)Q({\bf R},2\tau)\nonumber\\
 + \varepsilon&_3({\bf R},2\tau)G({\bf R},2\tau)-2  \varepsilon_4({\bf R},2\tau)F({\bf R},2\tau)\big]\, ,
\label{eq:secorder}
\end{align}
where $S_{\bf k}$ and $G_{\bf k}$ are given by Eqs. (\ref{eq:sk}) and (\ref{eq:gk}), respectively, and
$P({\bf R}):=P({\bf R},0)$, and so on. $A_1$, $A_2$, $C_1$ and $C_2$ are simple sums, but also in $B_1$ and $B_2$ there are no true double sums because the ${\bf k}$- and ${\bf q}$-dependent terms can be handled independently.

For periodic boundary conditions, where $\varepsilon_{\bf k}$ is even under a reflection by the diagonal and $\Delta_{\bf k}$ is odd, the above expressions are simplified. We have chosen periodic-antiperiodic boundary conditions, for which this symmetry does not hold as long as $L$ remains finite. Therefore we have used the full expressions (\ref{eq:secorder}) in the computations.
\section{BCS pairing for nearest-neighbor attraction}
\label{B}
The extended Hubbard model with repulsive on-site and attractive nearest-neighbor interactions, defined by the Hamiltonian
\begin{align}
H=\sum_{\bf k\sigma}\varepsilon_{\bf k}c_{\bf k\sigma}^\dag c_{\bf k\sigma}^{\phantom{}}+U\sum_{\bf R}n_{\bf R\uparrow}n_{\bf R\downarrow}\nonumber\\
\qquad-V\sum_{\langle{\bf R,R'}\rangle}n_{\bf R}n_{\bf R'}
\end{align}
($U>0$, $V>0$, $n_{\bf R}:=\sum_\sigma n_{\bf R\sigma}$), is unstable with respect to $d$-wave pairing at the BCS mean-field level. Using the ansatz of Section \ref{41}, one readily finds the expectation value
\begin{align}
\frac{1}{N_s}\langle H\rangle&=\frac{2}{N_s}\sum_{\bf k}\varepsilon_{\bf k}P_{\bf k}+U\left(\frac{n^2}{4}+f_0^2\right)\nonumber\\
& -2V\left\{n^2-\sum_{\alpha=x,y}\left[\left(\frac{1}{N_s}\sum_{\bf k}\cos k_\alpha P_{\bf k}\right)^2\right.\right.\nonumber\\
&\left.\left.-\left(\frac{1}{N_s}\sum_{\bf k}\cos k_\alpha F_{\bf k}\right)^2\right]\right\}\, ,
\label{eq:energy_bcs}
\end{align}
where $P_{\bf k}$, $F_{\bf k}$ and $f_0$ are defined, respectively, by Eqs. (\ref{eq:pk}), (\ref{eq:fk}) and (\ref{eq:f0}).
\begin{table}
\begin{tabular}{lllll}
\, n&\, V&\quad $\Delta_0$&\quad $\varepsilon_c$&$\varepsilon_c/(N(0)\Delta_0^2)$\\
\hline
0.8&\, 0.5&\, 0.010131&\, $1.541\times 10^{-5}$&\, 0.8159\\
0.8&\, 0.8&\, 0.089255&\, $1.187\times 10^{-3}$&\, 0.8096\\
0.8&\, 1.0&\, 0.176589&\, $4.555\times 10^{-3}$&\, 0.7939\\
0.6&\, 0.5&\, 0.000099&\, $5.936\times 10^{-10}$&\, 0.4376\\
0.6&\, 0.8&\, 0.009960&\, $6.738\times 10^{-6}$&\, 0.4881\\
0.6&\, 1.0&\, 0.047185&\, $1.480\times 10^{-4}$&\, 0.4777\\
\end{tabular}
\caption{Gap parameter $\Delta_0$ and condensation energy $\varepsilon_c$ for different densities $n$ and interaction strengths $V$, as obtained by minimizing the energy (\ref{eq:energy_bcs}) for $L=1000$. To get the ratio $\varepsilon_c/(N(0)\Delta_0^2)$, one also needs the density of states at the Fermi energy, $N(0)$. We find $N(0)\approx 0.18399$ for $n=0.8$ and $N(0)\approx 0.13912$ for $n=0.6$ in the thermodynamic limit, $L\rightarrow\infty.$ }
\label{table:BCS}
\end{table}
The minimization of this expression with respect to the gap parameter $\Delta_0$ for a fixed density $n$ (this constraint determines the chemical potential $\mu$) gives equilibrium values for $\Delta_0$ and for the condensation energy $\varepsilon_c$, as shown in Table \ref{table:BCS}. Together with the density of states $N(0)$ these data allow us to find the combination $\varepsilon_c/(N(0)\Delta_0^2)$, which is just $\frac{1}{2}$ in the original BCS theory for $s$-wave pairing. The corresponding numbers of Table \ref{table:BCS} for $d$-wave pairing are all quite close to $\frac{1}{2}$ although gap parameters and condensation energies vary by several orders of magnitude.

\section{Second-order terms for the order parameter}
\label{C}
The second-order terms of the expansion (\ref{eq:expop}) correspond to the diagrams of Fig. \ref{fig:second_order}, and we write
\begin{align}
\langle(C_{\bf k}^\dag+C_{\bf k}^{\phantom{}})D^2(\tau)\rangle_c&=A_1+B_1+C_1\, ,\nonumber\\
\langle D(-h)(C_{\bf k}^\dag+C_{\bf k}^{\phantom{}})D(\tau)\rangle_c&=A_2+B_2+C_2\, .
\end{align}
Using the notations of Sections \ref{41} and \ref{42}, we find
\begin{align}
A_1&=e^{-2\tau E_{\bf k}}G_{\bf k}S_{\bf k}(n G_{\bf k}-4f_0F_{\bf k})\, ,\nonumber\\
B_1&=e^{-2\tau E_{\bf k}}G_{\bf k}
\frac{1}{N_s}\sum_{\bf q}S_{\bf q}(G_{\bf k}G_{\bf q}+4F_{\bf k}F_{\bf q})\, ,\nonumber\\
C_1&=2e^{-2\tau E_{\bf k}}G_{\bf k}\sum_{\bf R}\cos {\bf k\cdot R}\nonumber\\
&[F^2({\bf R})-P({\bf R})Q({\bf R})][G_{\bf k}F({\bf R})-F_{\bf k}G({\bf R})]\, ,
\label{eq:second1}
\end{align}
and
\begin{align}
A_2&=-4e^{-2\tau E_{\bf k}}F_{\bf k}S_{\bf k}^2\, ,\nonumber\\
B_2&=e^{-4\tau E_{\bf k}}G_{\bf k}\frac{2}{N_s}\sum_{\bf q}e^{-4\tau E_{\bf q}}S_{\bf q}\nonumber\\
&\qquad\qquad(-Q_{\bf k}P_{\bf q}-P_{\bf k}Q_{\bf q}+2F_{\bf k}F_{\bf q})\, ,\nonumber\\
C_2&=4e^{-2\tau E_{\bf k}}F_{\bf k}\sum_{\bf R}\cos {\bf k\cdot R}\nonumber\\
&[F^2({\bf R},2\tau)-P({\bf R},2\tau)Q({\bf R},2\tau)]\nonumber\\
&[P_{\bf k}Q({\bf R},2\tau)+Q_{\bf k}P({\bf R},2\tau)-2F_{\bf k}F({\bf R},2\tau)]\, .
\label{eq:second2}
\end{align}

\end{appendix}
\bibliography{Citations.bib}
\end{document}